\documentclass[a4paper,11pt]{article}
\usepackage{aaskaiid}
\usepackage{hyperref}
\usepackage{xspace}
\usepackage{orcidlink}

\newcommand{\arcsec}{\ensuremath{^{\prime\prime}}\xspace}
\newcommand{\farcs}{.\!\!^{\prime\prime}}

\title{Cradle of Life: From the Formation of Stars to Habitable Worlds with the SKAO}
\ShortTitle{Cradle of Life}

\author[1]{Eleonora Bianchi\orcidlink{0000-0001-9249-7082}}
\ShortName{Bianchi \& Callingham} % shortened name list for header 
\author[2,3]{Joseph R. Callingham\orcidlink{0000-0002-7167-1819}}
\author[]{the Cradle of Life working group}

\affiliation[1]{INAF, Osservatorio Astrofisico di Arcetri, Largo E. Fermi 5, I-50125, Firenze, Italy}
\emailAdd{eleonora.bianchi@inaf.it; callingham@astron.nl}

\affiliation[2]{ASTRON, Netherlands Institute for Radio Astronomy, Dwingeloo, The Netherlands}

\affiliation[3]{Anton Pannekoek Institute for Astronomy, University of Amsterdam, Amsterdam, The Netherlands}

\abstract{This chapter provides an overview of the different science cases covered by the "Cradle of Life" working group, which aims to leverage the capabilities of the Square Kilometre Array Observatory (SKAO) to trace the physical and chemical pathways toward stars and planets formation, how such stars impact their planets, and whether life could exist in such conditions. At its highest frequencies, the SKAO will probe the earliest stages of the raw material fuelling planet formation, while enabling deep, unprecedented searches for prebiotic molecules both in high-mass and solar-type protostars. Concurrently, the lowest frequencies will be deployed to detect and characterize exoplanetary magnetic fields via their auroral radio emission, and the type of space weather a planet experiences from its host star. Across the entire frequency range, the SKA telescopes will conduct systematic searches of technosignatures. Together, these high-impact research areas establish a comprehensive roadmap for the SKAO to uncover the origins of life in the universe.}

\begin{document}

\maketitle

\section{Introduction}
%\eb{Total lenght 5 to 10 pages MAX. Example of citation is \cite{Garufi2026}. A concise overview of the scope and aims of the science pursued by the SWG.}\\

The fundamental science goals of the Cradle of Life working group follows a physically interlinked, temporal sequence: we want to understand how stars and planets formed, how such stars impact their planets, what characteristics these planets have, and, ultimately, whether signatures of life could be discoverable from such stellar systems. We are at an interesting point in the Cradle of Life working group as the established facilities, such as the Atacama Large Millimeter/submillimeter Array (\href{https://almascience.eso.org/}{ALMA}), and SKA precursors/pathfinders, have made impressive inroads in these science areas \citep[e.g.,][]{Callingham2024, Tobin2024, Ceccarelli2023, Motte2018, Cordiner2025}. What the SKAO stands to contribute in this field will be complementary to what has recently emerged in the literature, and ground breaking in many others. 

As an example, a combination of facilities, including space-based observatories like the James Webb Space Telescope \href{https://www.esa.int/Science_Exploration/Space_Science/Webb}{(JWST)}, and ground-based interferometers such as \href{https://almascience.eso.org/}{ALMA}, 
%the Northern Extended Millimetre Array \href{https://iram-institute.org/observatories/noema/}{(NOEMA)}, and the Very Large Array \href{https://public.nrao.edu/telescopes/vla/}{(VLA)}, and single dishes like the \href{https://iram-institute.org/observatories/30-meter-telescope/}{IRAM 30-m}, the Green Bank Telescope \href{https://public.nrao.edu/telescopes/gbt/}{(GBT)}, and the \href{https://rt40m.oan.es/}{Yebes 40-m}, 
has significantly advanced our understanding of the stars and planets formation process as well as on the origin of our Solar System \citep[see e.g.,][]{Radley2025}. Unlocking a new spectral window, the SKA-low (50--350 MHz) and SKA-Mid (0.35--15.4 GHz) telescopes will reveal previously hidden components of the gas and dust involved in planet formation. Specifically, they will allow us to probe centimetre-sized dust grains, the crucial pebbles from which planets begin to form in disks \citep{Ilee2020}. Furthermore, SKA-Mid will serve as an extraordinary detector of large prebiotic molecular species in star-forming regions \citep{Jimenez2022}.

Another example is the population of radio bright stars which is just beginning to be revealed by such telescopes as \href{https://www.csiro.au/askap}{ASKAP} and \href{https://www.astron.nl/telescopes/lofar/}{LOFAR}, but we are likely just seeing the tip of the iceberg. It is also currently unclear if a part of this radio bright star population is radio bright due to the presence of a nearby exoplanet. With the unprecedented sensitivity at both gigahertz and megahertz frequencies, SKAO will undoubtedly reveal the rest of this radio bright stellar system iceberg, potentially also establishing whether radio astronomy is ready to become an exoplanet discovery and characterization tool. 

To reflect the aforementioned interlinked physical sequence, we have arranged this overview chapter to reflect the natural evolution in time of the chapters discussed here -- starting with what we will discover with the SKAO about star and planet formation to the radio characteristics of those stars and planets, and finally, to whether life could be present on such planets. The topics covered by the Cradle of Life working group are highly complementary to those of Our Galaxy, which focuses specifically on the study of the Milky Way. For this reason, in addition to the chapters developed by Cradle of Life, we also present chapters from Our Galaxy that are relevant to each scientific area.

\section{Stars and planets formation}

\begin{figure}[h]
    \centering
	\includegraphics[width=1\columnwidth]{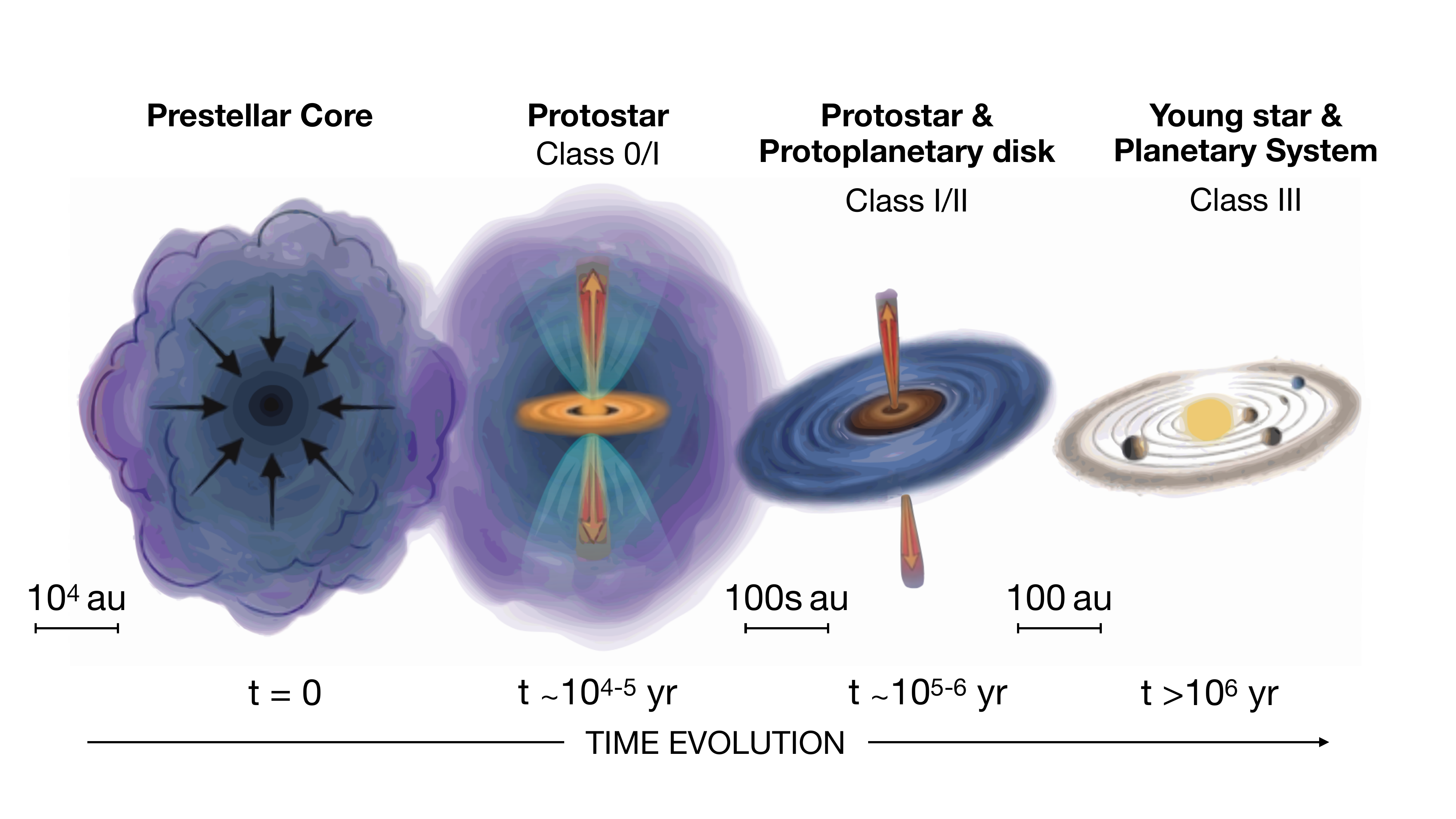}
    % Note the backslash (space) after \footnotemark
    \caption{The figure illustrates the four main phases of low-mass star formation: from the gravitational collapse of the natal core of gas and dust (Prestellar Core), to the main accretion phase (Class 0/I protostar, age $\sim$10$^4$-10$^5$ yr), followed by the protostar and its surrounding protoplanetary disk (Class I/II, age $\sim$10$^5$-10$^6$ yr), and finally the revealed planetary system after the disk has dispersed (Class III, age $\sim$10$^7$ yr). Typical spatial and timescales are indicated for each stage. Generated with Google Gemini 3 Flash and modified.}
    \label{fig:sketch-star-formation}
\end{figure}

The process of star and planet formation begins when a cold ($\sim$10 K) and dense ($>$10$^4$ cm$^{-3}$) cloud undergoes gravitational instability and starts to collapse under its own gravity (e.g., \citealt{Shu1987,Stahler-Palla2004}). During the collapse, material accreting from the large-scale envelope toward the central protostellar embryo conserves its angular momentum, naturally leading to the formation of a protostellar disk \citep{Bate2018, Draz2023,Lebreuilly2024}. Simultaneously, to remove angular momentum a fraction of the infalling material is ejected by the system, driving bipolar collimated high-velocity ($\sim$ hundreds of km s$^{-1}$) jets and large-scale low-velocity ($\sim$ tens of km s$^{-1}$) outflows \citep{Frank2014}. Initially, the nascent star is deeply embedded within its parental envelope (Class 0/I sources, age $\sim$10$^4$-10$^5$ years, see e.g., \citealt{Lada1987, Andre1993}). This envelope is gradually accreted onto the central object or dispersed by outflows (Class I/II sources, age $\sim$10$^5$-10$^6$ years), until the newly formed star and its disk are finally revealed (Class III sources, age $\sim$10$^7$ yr). Figure \ref{fig:sketch-star-formation} provides a simplified schematic representation of the standard paradigm for low-mass star formation, showing its four primary evolutionary stages along with their typical spatial and temporal scales.

While this evolutionary sequence represents the standard paradigm for Sun-like stars, high-mass star formation (M $\gtrsim$ 8 M$_{\odot}$) is distinguished by a significantly higher radiation field due to higher protostellar luminosities (L $\gtrsim$ 10$^{3}$ L$_{\odot}$) and a consequently more intense radiative and mechanical feedback \citep{Motte2018,Beuther2025}. In this regime, the protostellar embryo ionizes the gas of its surrounding envelope, creating an HII region that expands within the cloud from sub-parsec scales to larger volumes. 

Planet formation occurs in the inner protostellar regions within the disks surrounding protostars, where dust grains grow and coagulate to form the rocky cores of planets, while the gas in the disk is accreted to form both the gas giants and the planetary atmospheres, in a process lasting approximately 10 million years (Myr) \citep{Testi2014,Draz2023, Birnstiel2024}. To connect the star and planet formation process to the final composition of planets, it is essential to characterize the physical structure and chemical composition of the star-forming regions across all scales, from the parental molecular clouds and infalling envelopes to the protoplanetary disks themselves \citep[e.g.,][]{Oberg2021}. Finally, the study of Solar System bodies such as asteroids, meteorites, and comets provides a unique opportunity to bridge the gap between other star forming regions and the specific formation history of our own Solar System \citep[e.g.,][]{B-Morvan2008,Lippi2024,Morbidelli2024}.

%Despite a well-established framework, and the substantial advancements thanks to facilities such as the Atacama Large Millimeter/submillimeter Array (ALMA \footnote{\href{ALMA}{www.almaobservatory.org/en/home/}}), the Karl G. Jansky Very Large Array (VLA\footnote{\href{VLA}{https://public.nrao.edu/telescopes/vla/}}), and the James Webb Space Telescope (JWST\footnote{\href{JWST}{https://science.nasa.gov/mission/webb/}}), many aspects of star formation still need to be understood. In this respect, the advent of the Square Kilometre Array Observatory (SKAO\footnote{\href{SKAO}{https://www.skao.int/en}}) will open a new frontier.

\subsection{Advancing our picture of star formation with the SKA telescopes}

The SKA telescopes will substantially advance our understanding of the star formation process, beginning with its earliest stages. 
The proposed SKA Galactic Plane survey, would provide a panoramic view of ionized gas and stellar feedback across the Milky Way at unprecedented sensitivity and angular resolution \citep{Traficante01.2026.SKA}. Deep, wide-field observations of HI and light molecules such as hydroxyl radical (OH), methylidyne radical (CH), and formaldehyde (H$_2$CO), will probe the multi-phase interstellar medium from its ionized to atomic and molecular phases, probing the star formation sites within our galaxy \citep{Yamamoto01.2026.SKA,Karska01.2026.SKA}. 
Furthermore, SKA-Mid will observe OH, methanol (CH$_3$OH), and H$_2$CO masers using bands 2, 5a, and 5b, followed by CH masers in band 4 \citep{Rygl01.2026.SKA}. Thanks to their compactness and high brightness temperature, these masers serve as precise "cosmic rulers" for astrometry, enabling the measurement of positions, structures, and kinematics in dense star forming regions that remain largely inaccessible at other wavelengths. These observations allow us to use the physical conditions found across the Milky Way as a template for understanding the processes driving star formation and galaxy evolution on a global scale. 
By utilizing the combined capabilities of SKA-Low and SKA-Mid, we will be able to detect and map magnetic fields within both high- and low-mass cores inside molecular clouds. Recent studies indicate that cosmic-ray electrons interacting with these magnetic fields can produce detectable synchrotron emission at low radio frequencies. This provides a novel and powerful probe of core magnetization, offering a window into the magnetic properties that dictate the dynamics of gravitational collapse and the subsequent birth of stars and their planetary systems \citep{Bracco01.2026.SKA}. Magnetic fields in molecular clouds and their role in the star formation process will also be studied via the Zeeman effect in spectral lines \citep{Bourke01.2026.SKA}.
As we evolve into the protostellar stage, SKA-Mid will allow us to probe the star formation feedback on the surrounding medium. A primary objective is the study of jets and outflows, where SKA-Mid will enable high-angular resolution observations at centimetre wavelengths \citep{Sabatini01.2026.SKA}. Radio recombination lines and proper motion measurements will allow us to reconstruct the 3D kinematics of protostellar jets while the detection of non-thermal, linearly polarized synchrotron emission will enable the measurement of magnetic field strengths and morphologies at unprecedented scales of just a few au, providing direct observational constraints on magnetohydrodynamic models of protostellar outflows.
Star formation studies will be further transformed through the strategic synergies between the SKAO and ALMA, which will yield significant results starting from the early science phase \citep{Forbrich01.2026.SKA}. This multi-wavelength framework addresses several critical frontiers in the field. First, simultaneous time-domain and variability studies of young stellar objects will open a new observing window to constrain the physics of extreme stellar flares and their interplay with mass accretion. Second, the two observatories provide essential complementary coverage of molecular transitions and radio recombination lines to probe the interstellar medium across a broad range of physical conditions. Finally, the synergy between the SKAO and ALMA will be significant, driven by both the ALMA2030 wideband sensitivity upgrade and the alignment in spatial resolution between the two observatories. This combination will provides a complete perspective of the star formation process across centimetre to sub-millimetre wavelengths.

 % Finally, the SKA-project will allow exploring the dust composition and chemical enrichment in shocks, where sputtering/shattering of grains cause the release of their mantles and refractory cores in the gas-phase. Complementary to ALMA’s detection of simple and complex organic molecules, the SKAO will probe, for the first time, long carbon chains/rings, several Cl-, Al-, Mg-, and other metal-bearing species (missed by current sub-mm facilities).

\subsection{Probing the first steps of planet formation}

SKA-Mid observations will be fundamental to studying how interstellar dust grains grow from micron size to form planetesimals and, eventually, planets. On the one hand, a first large-scale, high-resolution survey of disk emission at centimetre wavelengths, will provide a comprehensive census of pebbles within the planet-forming disks of nearby star-forming regions \citep{Garufi01.2026.SKA, Greaves01.2026.SKA}. By resolving the spatial distribution, spectral properties, and evolutionary trends of these pebbles, SKA-Mid observations will offer essential constraints on dust growth and disk dynamics. Furthermore, these observations will allow for a more accurate determination of total dust mass in disks and potentially uncover the presence of protoplanets and their associated circumplanetary disks.
On the other hand, SKA-Mid Band 5b continuum observations at very high-angular resolution (from 0$\farcs$05 to 0$\farcs$15) at 12.5 GHz, will reveal variations in density, temperature, and composition of the dusty disk shedding light on the origin of substructures such as rings, gaps, spirals, and vortices \citep{Wu01.2026.SKA}.
The study of grain growth also extends beyond the disk itself. Observations of dust emission within cavities created by the interaction of protostellar outflow with the surrounding medium, will allow for an investigation into how dust grows and is transported from the disk to the surrounding envelope \citep{Sabatini01.2026.SKA}.

Beyond the solid dust component, SKA-Mid observations will provide key constraints on planet formation models by probing the ionized disk component, which traces critical processes such as photoevaporation, accretion, disk winds, and jets \citep{Guidi01.2026.SKA}. State-of-the-art simulations of magneto-thermal and magnetohydrodynamic winds demonstrate the potential of the SKA telescopes to detect and characterize free-free emission and hydrogen recombination lines, casting light on the mechanisms driving disk evolution and the eventual onset of planet formation.

Finally, SKA-Mid observations will provide a unique probe into the origin of our own Solar System by characterizing the thermal emission of Trans-Neptunian Objects (TNOs) and Centaurs \citep{Santos-Sanz01.2026.SKA}. These distant, icy bodies are essential for understanding the evolution of the outer Solar System and for establishing vital links with the protoplanetary and debris disks observed around other stars. 
The SKA telescopes will also open a new observational window through radio occultations, enabling independent measurements of object sizes, shapes, atmospheres, rings, and satellites. Combined with observations from ALMA, JWST, and traditional stellar occultation campaigns, these capabilities will establish centimetre-wavelength astronomy as a powerful tool for exploring the thermal and structural properties of outer Solar System bodies.

\subsection{Unveiling Chemical Complexity with the SKAO}

SKA-Mid is set to revolutionize the field of astrochemistry, offering an unprecedented window into the chemical origins of life. While a high level of molecular diversity has been observed in the early stages of star formation, SKA-Mid will fundamentally expand this inventory by uncovering large, heavy-atom molecules that possess rotational transitions at radio frequencies \citep{Jimenez2022, Bianchi01.2026.SKA}. SKA-Mid Band 5 will enable to investigate for the first time the densest protostellar regions, which remain heavily obscured by dust opacity at shorter wavelengths.
The extraordinary sensitivity of SKA-Mid especially in Band 5b, bolstered by the integration of the MeerKAT antennas, will push the boundaries of detection of complex species, enabling us to identify molecular species with fractional abundances as low as $\sim$10$^{-12}$ relative to H$_2$.
This will allow us to map long carbon chains and rings, enabling an investigation into the origin of the observed chemical diversity around young protostars within galactic star-forming regions. SKA-Mid will enable the exploration of deuterated isotopologues (molecules where deuterium replace one or more hydrogen atoms) and the chemistry of galactic environments at low metallicities.
For the first time, SKA-Mid will provide the sensitivity and angular resolution necessary to resolve the chemical composition of protoplanetary disks down to solar system scales ($\sim 30$ au; \citealt{Podio01.2026.SKA}). Crucially, we will be able to explore the disk midplane, a region impenetrable to shorter-wavelength observations due to high dust opacity. By constraining the initial conditions of disk evolution and planet formation, these observations will allow us to predict the chemical makeup of forming planets and their atmospheres. Comparing these findings with upcoming data on exoplanets' atmospheres and pristine Solar System bodies, such as comets and asteroids, will provide vital new insights into the origin and evolution of planetary systems, including our own \citep{Podio01.2026.SKA}.

While deep integrations, representing a Key Science Project of approximately 1 000 hours, are essential to fully probe the chemical complexity of these regions, shorter observation times ($\lesssim$ 10 hours) using the initial AA* array configuration will still enable early science with the detection of 
prominent species such as cyanoacetylene (HC$_3$N) and cyanodiacetylene (HC$_5$N) in nearby star forming regions. A clear observational advantage of SKA-Mid is the large Field of View, approximately 7$^{\prime}$ in Band 5b.
 For comparison, that of ALMA is between 23$\arcsec$ and 30$\arcsec$ in Band 6 (211-275 GHz). The SKA-Mid wide field
 will allow to target dozens of sources in regions such as Ophiuchus and Orion (d $\sim$ 140-400 pc) simultaneously. Compared to deep single-dish observations at equivalent frequencies, SKA-Mid provides the critical capability to resolve molecular emission on much smaller spatial scales (down to 100 au). This high angular resolution is fundamental to investigating chemical formation and destruction pathways, as well as mapping molecular distributions within the planet-forming region. Finally, the deep integrations required for molecular line observations offer high potential for commensality, as the resulting high-sensitivity data will simultaneously benefit continuum and maser studies.

\begin{figure}[h]
    \centering
	\includegraphics[width=1\columnwidth]{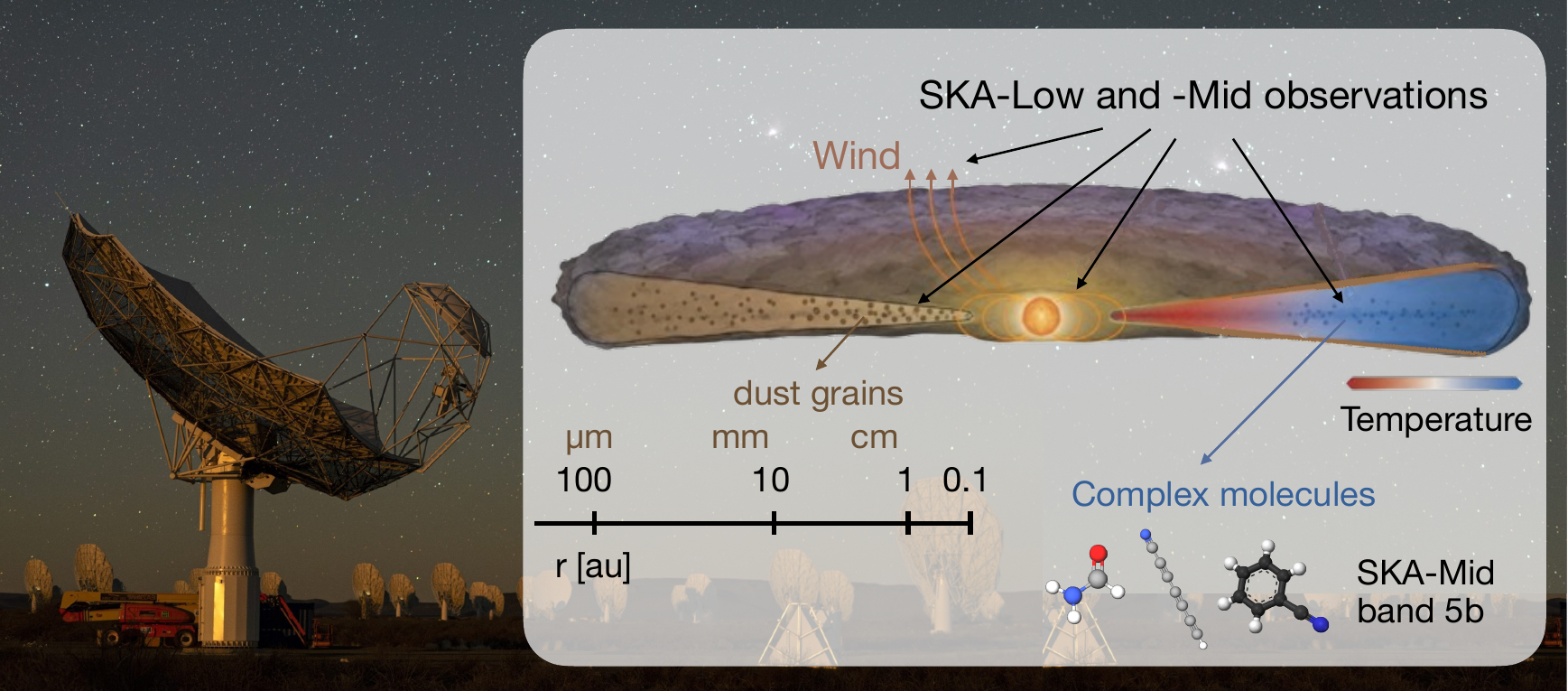}
    % Note the backslash (space) after \footnotemark
    \caption{This composite illustration depicts a sketch of a protoplanetary disk and its corresponding spatial scales. SKA-Low and SKA-Mid observations will probe critical physical processes, such as dust grain growth, dust and gas substructures, and ejection processes such as winds and jets. Moreover, SKA-Mid in band 5b will be able to detect species such as complex organic molecules, carbon chains and ring (as depicted in the illustration) emitted close to the disk midplane, the planet-formation region.
   Image credits: SKAO/Max Alexander (SKA-Mid antenna); artistic layout inspired by the illustration in the SKAO "Cradle of Life" poster by D. Dall' Olio, modified using Google Gemini and subsequently by the authors.}
    \label{fig:SKA-disks}
\end{figure}

\section{Stars, exoplanets, space weather and the search for life}
\label{sec:exoplanets-spi-seti}

The formation of stars and planets represents only the first stage in the emergence of potentially habitable worlds. Once planetary systems are assembled, their long-term evolution is governed by the interplay between stellar radiation, magnetic activity, winds, energetic particles, and planetary magnetospheres. These processes regulate atmospheric escape, surface and atmospheric irradiation, and the broader space-weather conditions under which life may emerge and persist. 

The SKAO, especially with its unparalleled sensitivity at low-frequencies, will uniquely connect these themes by probing stellar and planetary magnetic fields, detecting radio emission from exoplanets and ultracool dwarfs, characterizing magnetic star--planet interactions, and conducting sensitive searches for technosignatures across a wide range of astrophysical environments. The Chapter by \citet{Driessen01.2026.SKA} in the Transients science working group and those by \citet{Curiel01.2026.SKA, Cavallaro01.2026.SKA} in the Our Galaxy working group also touches on some of the areas discussed below. 

\subsection{Exoplanetary magnetospheres and radio emission}
\label{subsec:exoplanet-magnetospheres}

Magnetic fields are thought to play a fundamental role in regulating the atmospheric evolution of planets. They can shield planetary atmospheres from stellar winds, mediate atmospheric escape, and provide direct information on the internal structure and dynamo processes of planets. However, despite the discovery of thousands of exoplanets, direct measurements of exoplanetary magnetic fields remain largely inaccessible through conventional observing techniques. Low-frequency radio observations provide a unique solution to this problem.

As discussed in the companion chapter by \citet{Kavanagh01.2026.SKA}, magnetized planets in the Solar System produce bright radio emission generated by energetic electrons interacting with planetary magnetic fields. This auroral emission, powered by the electron cyclotron maser instability, is highly polarized, strongly beamed, and emitted close to the local cyclotron frequency. Its detection therefore provides a direct diagnostic of the magnetic field strength and plasma conditions at the emission site. While a conclusive detection of radio emission from an exoplanet is still lacking, the unprecedented sensitivity of SKA-Low is expected to enable the first robust detections of magnetized giant exoplanets, opening a new observational window on planetary magnetic fields, atmospheric evolution, and habitability. This is an information that is impossible to obtain, or heavily model dependant, at other wavelengths.

The same SKAO observations will also transform the study of ultracool dwarfs. These objects have radii comparable to Jupiter but significantly larger masses, and their radio emission provides an important bridge between stellar and planetary magnetism. As \citet{Kavanagh01.2026.SKA} details, large SKAO surveys are expected to detect thousands of radio-emitting ultracool dwarfs within a few hundred parsecs, enabling statistical studies of magnetic activity at planetary scales. In combination with very long baseline interferometry, SKAO astrometry will also enable searches for planetary companions around nearby radio-emitting ultracool dwarfs through their reflex motion, complementing transit and radial-velocity surveys \citep{Kavanagh01.2026.SKA}.

\subsection{Stellar activity and planetary space weather}
\label{subsec:space-weather}

Planetary environments are continuously shaped by the activity of their host stars. Stellar winds, flares, coronal mass ejections, and magnetic reconnection events inject energetic particles into circumstellar environments and can significantly alter planetary atmospheres over gigayear timescales. These effects are particularly important for planets orbiting low-mass stars, where habitable-zone planets are located close to their hosts and are therefore exposed to intense magnetic activity and strong space-weather forcing.

The space-weather environment of an exoplanet determines the rate at which its atmosphere is eroded, heated, or chemically modified. Stellar winds and coronal mass ejections can strip atmospheres, while time-variable magnetic fields can induce currents and heating in planetary ionospheres. Characterizing these processes requires observational constraints on stellar magnetic fields, energetic particles, wind conditions, and the response of planetary magnetospheres. As emphasized by \citet{Vedantham01.2026.SKA}, radio observations provide a direct probe of these magnetic and plasma processes. The SKAO, with its sensitivity and time-domain capabilities, will therefore play a central role in measuring stellar magnetic activity and quantifying its impact on planetary habitability.

\subsection{Magnetic star--planet interactions}
\label{subsec:spi}

A particularly direct manifestation of the coupling between stars and planets is magnetic star--planet interaction. In close-in planetary systems, a planet moving through the magnetized plasma environment of its host star acts as an obstacle to the stellar wind. Depending on whether the relative plasma flow is sub- or super-Alfv\'enic, the intercepted energy may be dissipated locally in the planetary magnetosphere, transported back toward the star along Alfv\'en wings, or both. This interaction can accelerate charged particles, generate auroral emission, heat stellar chromospheres and coronae, and potentially enhance stellar flare activity \citep{Vedantham01.2026.SKA}.

Radio emission from magnetic star--planet interactions is expected to be produced by the electron cyclotron maser instability. This emission can arise either from the planetary magnetosphere or from the stellar magnetic environment, depending on the interaction regime and plasma conditions. Since the maximum emission frequency is set by the local cyclotron frequency, radio detections provide a direct measurement of the magnetic field strength of the emitting body. This makes radio observations uniquely powerful compared with indirect optical, ultraviolet, or X-ray activity tracers.

Recent observations at optical, ultraviolet, and X-ray wavelengths have provided growing evidence for magnetic star--planet interactions in close-in planetary systems. However, a definitive radio detection remains elusive, primarily because previous facilities lacked the required sensitivity and because long-term monitoring of promising systems has been limited. The SKAO will transform this field by enabling systematic searches for radio signatures of magnetic star--planet interactions in nearby systems. Such observations will constrain planetary magnetic fields, stellar wind properties, interaction energetics, and atmospheric escape, establishing radio observations as a key component of exoplanet habitability studies \citep{Vedantham01.2026.SKA,Kavanagh01.2026.SKA}.

\subsection{Technosignatures}
\label{subsec:seti}

The final stage of the Cradle of Life science pathway is the search for life itself, including the possibility that life elsewhere has developed technology. As discussed by \citet{Tremblay01.2026.SKA}, technosignature searches have expanded rapidly over the last decade and are increasingly integrated into mainstream astrophysics.

Radio wavelengths remain particularly attractive for technosignature searches. Radio signals propagate efficiently through interstellar space, can be generated with relatively modest energy requirements, and may be distinguishable from natural astrophysical sources through their spectral, temporal, polarization, or modulation properties. Modern searches are no longer restricted to traditional narrow-band beacons, but also include broadband signals, drifting carriers, pulsed and transient emission, scintillating signals, and searches over both Galactic and extragalactic scales.

The SKAO will provide an unprecedented platform for these searches. Its combination of sensitivity, field of view, angular resolution, spectral resolution, and computational infrastructure will enable technosignature surveys over a vastly expanded volume of parameter space. Commensal observing modes will allow technosignature searches to be carried out alongside other SKAO science programmes, while machine-learning classification and multi-wavelength follow-up will be essential for distinguishing terrestrial radio-frequency interference, natural astrophysical phenomena, and credible candidates. As emphasized by \citet{Tremblay01.2026.SKA}, the SKAO will therefore be one of the most powerful facilities ever built for testing the prevalence of technological civilizations in the Universe.

\section{Conclusion}

Together, the science cases described in the companion chapters of the Cradle of Life science working group form the Cradle of Life roadmap. 
%They fully explore the sequence from the formation of stars, disks, planets, and chemical complexity, to the characterization of planetary environments, the assessment of habitability, and the search for technological life. 
By linking the formation of stars and planets to stellar magnetic activity, exoplanetary magnetospheres, space weather, magnetic star--planet interactions, and technosignatures, the SKAO will provide a comprehensive observational framework for addressing one of the most fundamental questions in astrophysics: can life exist outside of the Solar System? The science cases presented by the Cradle of Life working group emphasize the necessity of full baselines for SKA-Mid AA4, where high angular resolution is critical, particularly for studying protoplanetary disks.

At the time of writing, the Cradle of Life scientific community has grown to approximately 250 members spanning more than 20 countries. Beyond conducting cutting-edge science with SKA precursors and pathfinders, the working group is actively coordinating internal efforts to prepare for early science operations and the strategic planning of Key Science Projects. 
Given the strong public interest in Cradle of Life research, this working group will significantly contribute to engaging the general public and showcasing the broader impact of the SKAO. Furthermore, the science cases developed by the Cradle of Life community will help drive future observatory upgrades, such as the implementation of SKA-Mid Band 6, the future developement of SKA-Mid at higher frequencies (up to 50 GHz; \href{https://www.skao.int/sites/default/files/documents/d38-ScienceCase\_band6\_Feb2020.pdf}{SKA1 Beyond 15GHz}).

%Up to the writing of this book the scientific community of Cradle of Life has grown to about 250 members from more than 20 different countries. The working group is doing not only scientific work with SKAO precursors and pathfinders but also an effort of internal coordination to be ready for the early science and then the planning of the Key Scince projects with the SKA telescopes.
%Given the strong public interest in Cradle of Life research topics,
%this working group will significantly contribute in sharing the impact of the SKAO.
%Finally some of science cases developed by Cradle of Life will highly benefit to a future developement of SKA-Mid at higher frequencies, the proposed Band 6 (up to 50 GHz, link to Memorandum https://www.skao.int/sites/default/files/documents/d38-ScienceCase_band6_Feb2020.pdf) 

\section*{Acknowledgments}
We thank the previous chairs of the Cradle of Life science working group, in particular: Cherry Ng, John Ilee, Laurent Lamy, Josep Miquel Girart, Izaskun Jiménez-Serra, Doug Johnstone, Andrew Siemion, Di Li and Melvin Hoare. We also thank the Cradle of Life core team for their assistance in coordinating the organization of this science book, as well as the Cradle of Life community for their contributions to the chapters. The full list of Cradle of Life members is available at this \href{https://www.skao.int/en/science-users/science-working-groups/105/cradle-life/407/cradle-life-science-working-group-members}{link}.

E.B. acknowledges support from the Italian Ministry for Universities and Research under the Italian Science Fund (FIS 2 Call – Ministerial Decree No. 1236 of 2023 August 1) grant FIS-2023-00170. J.R.C. acknowledges funding from the European Union via the European Research Council (ERC) grant Epaphus (project number 101166008) and from the Dutch research council (NWO) under the talent programme (Vidi grant VI.Vidi.243.136).

\bibliographystyle{abbrvnat-maxbibnames4.bst}
\bibliography{overview} % if your bibtex file is called example.bib

\end{document}